# Superconductivity in potassium-doped metallic polymorphs of MoS$_2$

Renyan Zhang[1], I-Ling Tsai[1], James Chapman[1], Ekaterina Khestanova[1], John Waters[2], Irina V. Grigorieva[1,*]

[1] School of Physics and Astronomy, University of Manchester, Oxford Road, Manchester M13 9PL, United Kingdom

[2] School of Earth, Atmospheric and Environmental Sciences, University of Manchester, Oxford Road, Manchester M13 9PL, United Kingdom

*Superconducting layered transition metal dichalcogenides (TMDs) stand out among other superconductors due to the tunable nature of the superconducting transition, coexistence with other collective electronic excitations (charge density waves) and strong intrinsic spin-orbit coupling. Molybdenum disulphide (MoS$_2$) is the most studied representative of this family of materials, especially since the recent demonstration of the possibility to tune its critical temperature, T$_c$, by electric-field doping. However, just one of its polymorphs, band-insulator 2H-MoS$_2$, has so far been explored for its potential to host superconductivity. We have investigated the possibility to induce superconductivity in metallic polytypes, 1T- and 1T'-MoS$_2$, by potassium (K) intercalation. We demonstrate that at doping levels significantly higher than that required to induce superconductivity in 2H-MoS$_2$, both 1T and 1T' phases become superconducting, with T$_c$ = 2.8 and 4.6K, respectively. Unusually, K intercalation in this case is responsible both for the structural and superconducting phase transitions. By adding new members to the family of superconducting TMDs our findings open the way to further manipulate and enhance the electronic properties of these technologically important materials.*

Low dimensionality is considered to be an important ingredient for high superconducting transition temperatures, as solid state systems with reduced dimensions are more prone to electronic instabilities resulting in the development of different types of electronic order. There has been growing interest in two-dimensional (2D) or layered materials that are naturally superconducting [1-4] or where a superconducting transition can be induced by intercalation [5-8], electrostatic doping [9-12] or pressure [13-16]. MoS$_2$ was one of the first TMDs where superconductivity was induced by intercalation with alkali- and alkali-earth metals [17-20], with critical temperatures T$_c$ between 3K to 7K, depending on the intercalant [18,19]. More recently, it was found to become superconducting under ultrahigh pressure (T$_c$ =11.5 K at 90 GPa) [13] and a 'superconducting dome' (carrier-concentration dependent $T_c$ with a maximum of 10.8 K) was found at the surface of MoS$_2$ crystals under electrostatic doping [9,10]. However all the above studies were done on the semiconducting polymorph, 2H-MoS$_2$, while the potential for superconductivity in its metallic cousins remained unexplored, either theoretically or experimentally. Yet, they can be expected to exhibit properties different from other TMDs. For example, a competition between charge density waves (CDW) and superconductivity determines the behaviour of doped metallic 1T-type TMDs [21], including most recently studied TiSe$_2$ [22-24] and TaS$_2$ [15,25]. Such competition was also predicted for heavily electron-doped 2H-MoS$_2$ [26] and a question arises whether these electronic states also exist in metallic 1T-/1T'-MoS$_2$. Furthermore, electron doping was predicted to lead to unconventional superconductivity in MoS$_2$ [27,28] which is yet to be confirmed experimentally.

Polymorphism is a distinguishing feature of all layered transition metal dichalcogenides. In the case of MoS$_2$, depending on the relative arrangement of S and Mo atoms, the triple layers can form in one of three different configurations: trigonal prismatic (semiconducting 2H phase) [5,21], octahedral (metallic 1T phase) [29-30] and distorted octahedral with zigzag Mo-Mo chain structure (1T' phase) [31,32]. In addition to the coordination polytypism within the layers, there are several possibilities for the overall bulk symmetry, such as trigonal, hexagonal and rhombohedral [5,33]. As demonstrated recently, 2H to 1T/1T' phase transition in MoS$_2$ can be induced by heavy electron doping using Li [34-38] and Na intercalation [39,40], substitutional Re doping [41], electron-beam irradiation [30] and plasmonic hot electron doping [42]. It has also been predicted to be triggered by electron doping irrespective of the method used to achieve it, due to changes in the valence electron configuration of Mo atoms [43]. Importantly, while pure 2H phase is found in both naturally occurring and laboratory-grown MoS$_2$, 1T and 1T' phases usually coexist with the 2H phase in the same crystal, creating opportunities for phase engineering in electronic devices (e.g. ref. 35).

In this report, we demonstrate that potassium (K) intercalation induces 2H to 1T and 1T' phase transitions in MoS$_2$, similar to Li and Na, and that all three intercalated phases are superconducting, with $T_c \approx$ 6.9 K, 2.8 K and 4.6 K, respectively. The phase transitions are achieved by increasing K content in the samples. A continuing presence of K atoms between S-Mo-S layers is found to be essential for superconductivity: if K atoms are allowed to de-intercalate by exposure to air [18], superconductivity disappears, even though the structural changes caused by intercalation are expected to remain [35,38,44,45]. The constant $T_c$'s for all three superconducting phases indicate that intercalation leads to the formation of stoichiometric compounds not only for 2H-MoS2 (K$_{0.4}$MoS$_2$) as was established previously [18,20] but also for K-intercalated 1T- and 1T'-MoS$_2$.

To achieve K intercalation, we used the well-known liquid-ammonia method [5,17-19]. Briefly, intercalation was achieved by immersion of 2H-MoS$_2$ powder consisting of platelet-shape crystals (typical lateral size ~10 μm and thickness ~1 μm) in a solution of K metal in liquid ammonia at -78 °C (see Methods for details). K atoms in liquid ammonia are known to dissociate into solvated cations (K$^+$) and solvated electrons (e$^-$), forming a deep-blue solution [46]. As the solvated electrons are donated to empty Mo d-bands of 2H-MoS$_2$ and K$^+$ ions intercalate into van der Waals gaps to balance the charge, the solution gradually loses colour, which allows monitoring of the intercalation process. By varying the time of exposure to solvated K, we were able to obtain intercalated compounds with different average K concentrations. Furthermore, as demonstrated below, by using intercalation times up to 500 hours, we were able to achieve much higher degrees of intercalation (concentrations of K atoms) than previously thought to be possible [17-20].

Figure 1 shows the central result of our paper: K intercalation leads to the emergence of several superconducting phases. At intercalation times less than 40 hours, we found a broad but well defined superconducting transition at 6.9K - this is the same transition as found in early studies of K-intercalated MoS$_2$ and is known to correspond to a stoichiometric compound 2H-K$_{0.4}$MoS$_2$ with hexagonal symmetry [17-20]. As the intercalation time was increased further, two more superconducting transitions emerged, first at $T_c \approx$ 2.8K and then at $\approx$4.6K, neither of which was reported before. As follows from the susceptibility curves in Figs. 1 and 2a, either two or all three transitions were found in the samples intercalated for longer than 100 hours, indicating a coexistence of different superconducting phases. Here each $T_c$ was determined as a temperature ($T$) corresponding to a pronounced increase in diamagnetic susceptibility $\chi = M/H$ or, more quantitatively, as $T$ corresponding to a sharp increase in the derivative d$\chi$/d$T$, where $M$ is the magnetic moment and $H$ the applied magnetic field. Several examples are shown in Figs. 2a and Supplementary Fig. S1.

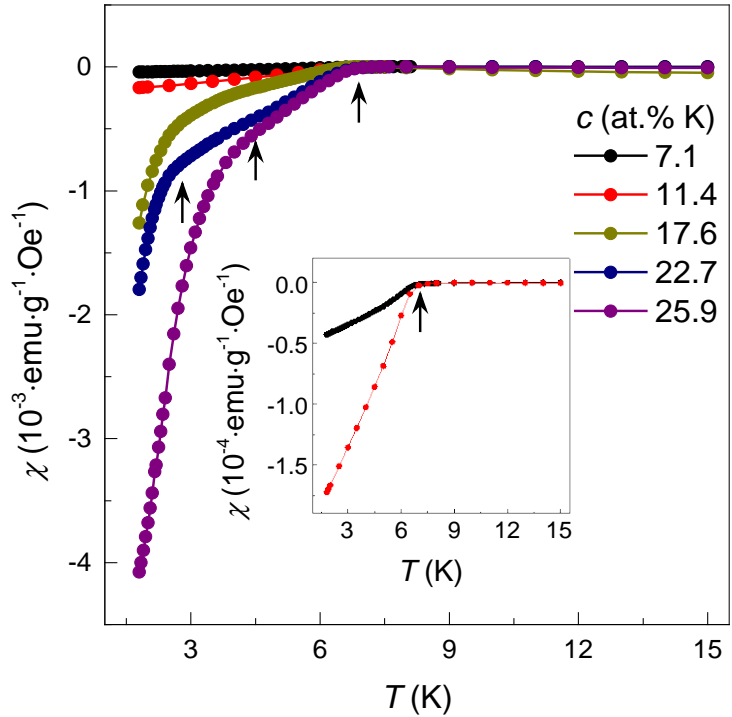

**Figure 1 | Superconductivity in K-intercalated MoS$_2$. Main panel:** Susceptibility $\chi = M/H$ for samples with increasing average concentration of potassium. The dc magnetic moment $M$ was measured in ZFC mode in an applied field $H$ =10 Oe. The average K concentrations, $c$, corresponding to different curves were determined using energy-dispersive X-ray microanalysis (see text). Arrows indicate the temperatures corresponding to the onsets of the three superconducting transitions as determined from d$\chi$/d$T$ in Fig. 2 and Supplementary Fig. S1. All three transitions are present for $c \approx$23 at.%. **Inset:** Zoom of the $\chi(T)$ curves for low K concentrations where only one superconducting transition is present. The higher $\chi$ for $c \approx$11 at.% corresponds to a larger superconducting fraction (lager intercalated volume in each MoS$_2$ crystallite) compared to $c\approx$7 at.%.

Although neither phase has completely replaced the others, the evolution of the susceptibility curves in Fig. 1 with intercalation time allowed us to conclude that the known 2H-K$_{0.4}$MoS$_2$ phase was the first to emerge at $T_c \approx$6.9K, followed by a second transition at $\approx$2.8K and then a third at $\approx$4.6K. Furthermore, comparison of the values of diamagnetic susceptibility for each transition showed that the fraction of the 6.9K phase increased monotonically up to the K concentration corresponding to the emergence of the new phases; the 2.8K phase only appeared at intermediate intercalations and was then replaced by the 4.6K phase which became dominant for the largest K content.

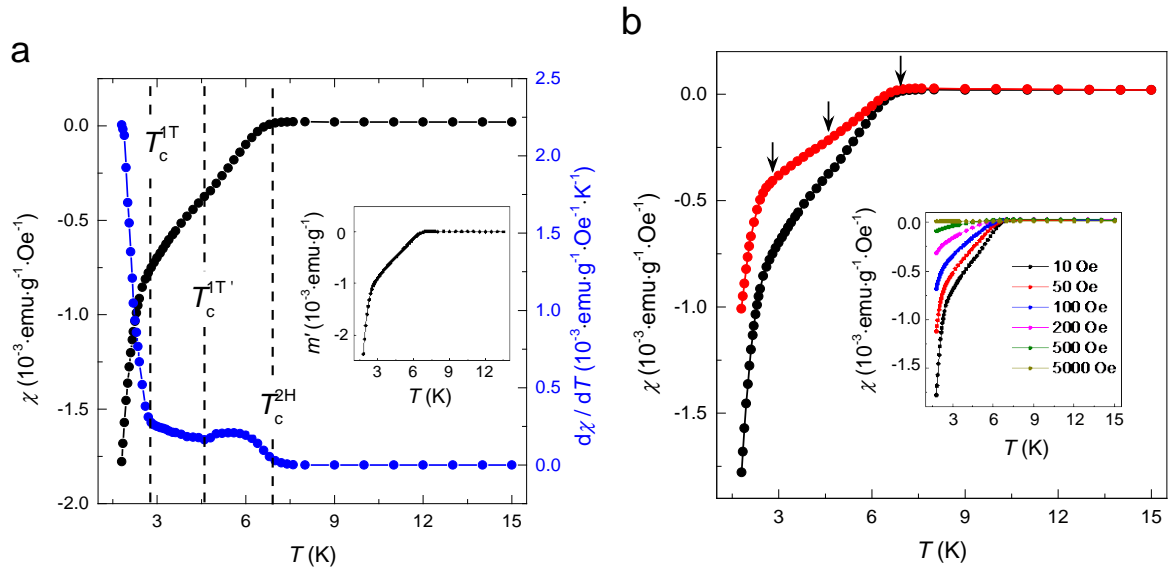

**Figure 2 | Superconducting transition temperatures and suppression of superconductivity by the magnetic field. (a)** dc susceptibility for $c \approx 23$ at.% K measured in ZFC mode at $H$ = 10 Oe and the corresponding numerical derivative d$\chi$/d$T$. Vertical dashed lines indicate the temperatures corresponding to the three superconducting transitions. The inset shows temperature dependence of a.c. susceptibility, $m'$, measured in zero applied field with an ac drive amplitude of 1 Oe and frequency 80 Hz. **(b)** Main panel: ZFC(black symbols) and FC (red symbols) susceptibility $\chi = M/H$ for $c \approx 23$ at.% K measured at $H$ = 10 Oe. The three superconducting transitions indicated by arrows are clearly seen both on ZFC and FC curves. Inset shows the evolution of $\chi$ with magnetic field.

To map the observed superconducting phases onto the samples' chemical composition and structural changes that are expected as a result of K intercalation [5], we used energy-dispersive X-ray spectroscopy (EDS), X-ray diffraction and X-ray photoelectron spectroscopy. The amount of K in the samples was determined using EDS (details in Methods, Supplementary Fig. S2 and Supplementary Table S1). This showed that, with increasing the intercalation time, the average K content increased from $\approx$ 4 at.% (intercalation for 24 hours) to $\approx$ 45 at.% (intercalation for 500 hours), i.e., we were able to achieve much higher K content than in earlier studies [18,19] where a well-defined stoichiometry, $K_{0.4}MoS_2$ ($\approx$ 12 at.%), was reported. Furthermore, as the elemental maps in Supplementary Fig. S2 show, the distribution of K after shorter intercalation times was non-uniform, with some crystallites showing higher K concentrations than others (Supplementary Fig. S2b). For longer intercalation times corresponding to the emergence of the new phases, this became essentially uniform (Supplementary Fig. S2c).

To understand whether the increasing K concentration in our experiments corresponds to a gradually increasing doping level, as in electrostatic gating experiments [9-12] or, rather, leads to one or more distinct crystal structures with different compositions, the samples – before and after intercalation - were characterised using X-ray powder diffraction (XRD) and X-ray photoelectron spectroscopy (XPS). Figure 3a shows the evolution of XRD spectra as the average K content, $c$, increases from $c \approx$ 4 at.% to $\approx$ 25 at.%. For K concentrations less than $\approx$12 at.% the most prominent (*002*) peak for pristine 2H-$MoS_2$, at $2\theta$ =14.4°, remains visible and a new peak at 10.7° appears and gradually increases in intensity. This new peak corresponds to an expanded lattice parameter along the *c*-axis (interlayer spacing) from 6.1(3) Å for pristine 2H-$MoS_2$ to 8.1(5) Å after intercalation. The in-plane lattice parameter remains almost unchanged, as the full refinement of the XRD spectra showed (peaks are labelled in Supplementary Fig. S3). Importantly, the observed peak shifts are not continuous with increasing K content, but rather correspond to a new crystal lattice with an increased interlayer spacing, as one would expect for K atoms filling the van der Waals gaps in MoS$_2$ crystals with a well-defined ratio of K to Mo atoms, i.e., a well-defined stoichiometric compound. Furthermore, the new lattice parameters match the known K$_{0.4}$MoS$_2$ compound with hexagonal structure studied in early intercalation experiments [17,18,20].

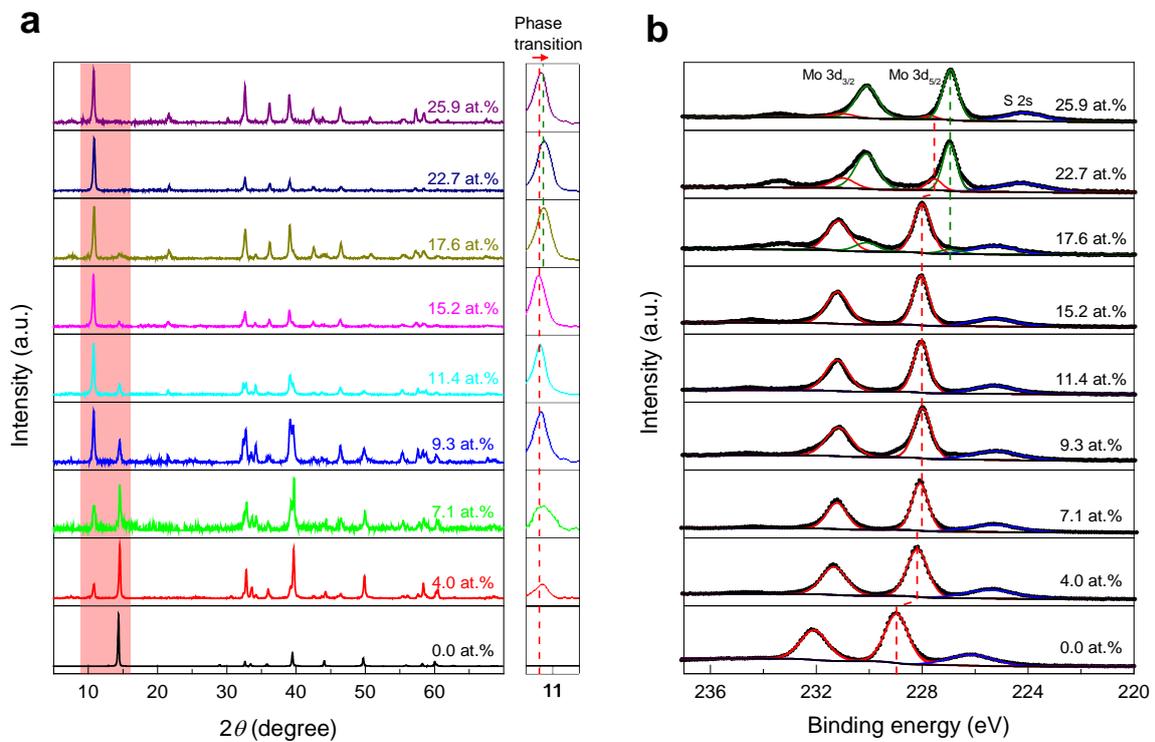

**Figure 3 | Characterization of pristine and K-intercalated MoS$_2$ samples. (a)** X-ray diffraction spectra for samples with gradually increasing K concentration. Numerical labels for each spectrum show the average K concentration obtained from EDS analysis. The spectral region corresponding to the position of (*002*) diffraction peak is highlighted to illustrate the emergence of an expanded crystal lattice due to the insertion of K atoms between S-Mo-S layers (see text). **Right panel:** Zoomed-up view of the (*002*) peaks corresponding to the lattice expansion after intercalation. A small shift to higher 2θ angle at *c*>18 at.% K indicates a 2H to 1T phase transition. **(b)** Evolution of high-resolution XPS spectra with gradually increasing concentration of K in intercalated samples. The shown energy region corresponds to the binding energies for Mo 3d and S 2s. Numerical labels show the average K concentration, same as in (a). Experimental spectra are shown in black; red and green curves correspond to Mo 3d and S2s peaks for 2H-K$_{0.4}$MoS$_2$ and 1T-KMoS$_2$ phases, respectively. The Mo 3d peaks first shift towards lower binding energies, then split into two sets of peaks corresponding to the coexisting 2H and 1T/1T' phases (see text).

Here we note that both the above techniques – XRD and EDS – provide data averaged over the volume of individual crystals, as the penetration depth of both X-rays and electrons significantly exceeds the typical crystal thickness of <1 µm. At the same time, the intercalation of K$^+$ starts at the surfaces of each crystallite and only gradually progresses toward the middle of the crystals as the intercalation time increases. Therefore the apparent continuous increase in the average K content from $c \approx 4$ at.% to $\approx 12$ at.% corresponds not to a changing stoichiometry but to an increasing fraction of each individual crystal's volume that becomes intercalated and turns into hexagonal K$_{0.4}$MoS$_2$. The intercalated fraction can be estimated from the ratio of integrated (*002*) peaks corresponding to pristine 2H-MoS$_2$ and K$_{0.4}$MoS$_2$. This yields that only ~20% of the crystals' volume are intercalated after 24 hours of exposure to solvated potassium (average K concentration ≈ 4 at.%) and this increases to 90% after~ 100 hours (measured K concentration ≈ 12 at.% corresponds to x≈0.4 in K$_x$MoS$_2$, or the stoichiometric compound K$_{0.4}$MoS$_2$).

As K content was increased further (measured K concentrations between ≈15 at.% and 26 at.%), we found clear signatures of a new phase transition: In XRD spectra it is seen as a small decrease of the interlayer spacing, with 2θ increasing from 10.7(5)° to 10.8(2)° (right panel of Figure 3a). A similar *c*-axis contraction was recently found for Na intercalation into 2H-MoS$_2$ in ref. 39 and was attributed to a transition from 2H to 1T phase. This transition is seen much clearer in X-ray photoelectron spectroscopy (XPS) measurements which probe the oxidation state of different constituent elements [47] – see Fig. 3b. The two peaks at 229 eV and 232 eV for Mo 3d$_{5/2}$ and Mo 3d$_{3/2}$, respectively, correspond to the +4 oxidation state of Mo in pristine 2H-MoS$_2$ [47]. After moderate intercalation (up to 12 at.% K or x=0.4 in K$_x$MoS$_2$) both peaks shift by ~1 eV, in agreement with earlier studies [45]. The lower binding energies after intercalation confirm that K donates electrons to MoS$_2$, resulting in a lower oxidation state of Mo [45,47]. We note that, unlike the gradual increase of the intensity of XRD peaks corresponding to K$_{0.4}$MoS$_2$ and their coexistence with pristine MoS$_2$, the above shift in binding energies is discontinuous, indicating that only the intercalated compound is present. The reason is that XPS probes only the top few nm of the crystals that are always fully intercalated (c.f. XRD/EDS discussion above).

The second phase transition seen in XRD spectra as a slight shift in the position of (002) peak, is much more obvious in XPS data. Here, deconvolution of the spectra for *c* >12 at.% reveals two new peaks downshifted in energy by ≈0.9eV. These first appear at ≈15 at.% K and becomes dominant at ≈25 at.%, that is, at K concentration corresponding to x≈ 1 in K$_x$MoS$_2$ (Fig. 3b). As shown in previous studies [35,44,45], the 0.9 eV shift in binding energy for Mo 3d$_{5/2}$ and Mo 3d$_{3/2}$ is a signature of a transition from the semiconducting 2H to the metallic 1T phase of MoS$_2$. The integrated intensity of the new peaks yields that the volume fraction of the 1T phase increases from ≈8% to ≈95% with the above increase of K content. The constant position of the new peaks indicates that the 1T phase induced by K intercalation is stoichiometric, similar to the 2H phase, and corresponds to a composition KMoS$_2$.

The accurate correspondence between the degree of intercalation required to achieve the structural transition to metallic 1T phase and the appearance of a new superconducting phase at 2.8 K is clear evidence that alkali-metal-doped metallic KMoS$_2$ is also a superconductor, albeit with a lower $T_c$ than its semiconducting counterpart. However, unlike the two new superconducting phases seen in magnetization measurements, the structural data seem to indicate only one phase transition. To this end, we recall that recent studies of heavily doped MoS$_2$ by transmission electron microscopy (TEM) reported that the octahedral 1T phase is only an intermediate between the trigonal 2H-MoS$_2$ and the more stable distorted octahedral phase (1T') where the octahedral 1T structure is distorted in such a way as to give chains of metal-metal-bonded Mo atoms [31,32,34,48]. Neither XRD, nor XPS measurements allow one to distinguish between these two metallic phases, because they correspond to the same interlayer separation and the same oxidation state of Mo. Nevertheless, as the 1T' phase has been shown to be more stable irrespective of the method used to induce the phase transition [31,32,34], the only possible explanation of our magnetization data is that the $T_c$=4.6K superconducting phase corresponds to the superconducting 1T' KMoS$_2$. The coexistence of superconducting 2H-K$_{0.4}$MoS$_2$ and either 1T or 1T' KMoS$_2$ is in agreement with the coexistence of these phases on 10-100 nm scale found in all TEM studies [30,32-34]. Following the latter studies, we expect that different superconducting phases are to be found in each individual crystallite, with their relative volumes changing as the average K concentration increases.

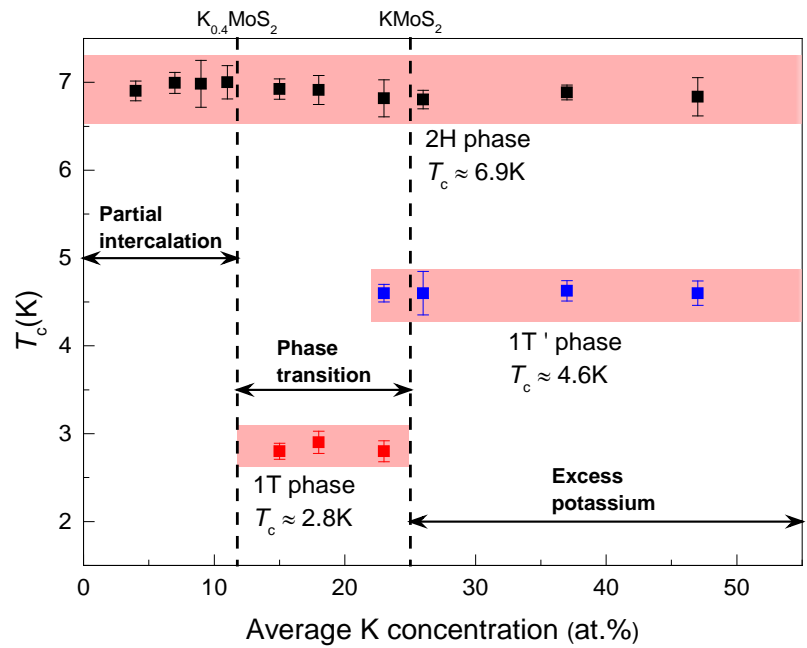

**Figure 4 | Superconducting phases of K-intercalated $MoS_2$.** Superconducting transition temperatures are shown as a function of average K content, *c*, obtained from EDS analysis. Error bars indicates the accuracy of determination of the onset transition temperatures. For K concentration *c* <12 at.%, $MoS_2$ crystals in our powder samples are only partially intercalated; accordingly, an increase in *c* corresponds to an increasing superconducting volume fraction of the 2H phase. For *c* >25 at.%, excess potassium causes a decomposition reaction and the volume fraction of both the 2H and 1T' phase starts to decrease but $T_c$ is not affected. Intermediate K concentrations, <12 at.% *c* <25 at.% correspond to an increasing volume fraction of the 1T phase.

We note that in all the above experiments it was important to protect the intercalated samples from exposure to oxygen and moisture, as it leads to de-intercalation of K atoms[18] and can be expected to affect both superconductivity and the structural transitions. To check this, we exposed some of our samples to air and measured the effect of exposure on magnetization and the oxidation state of Mo (XPS measurements). No superconducting response could be detected after exposure to air, while the XPS spectra indicated the continued presence of metallic 1T phase, i.e. intercalation-induced structural changes did remain (see Supplementary Fig. S4). Therefore, the continuing presence of K atoms is essential for superconductivity, similar to e.g. intercalated graphite[6].

As for higher still concentrations of K that we were able to achieve in our experiments (>25 at.%), instead of increasing the doping level, these were found to lead to a decomposition reaction where K started to bind to the S atoms in $MoS_2$ and produce $K_2S$ and metallic Mo - see Supplementary Fig. S5. In XRD spectra this is seen as a gradual decrease in intensity of the peaks corresponding to 1T/1T' $KMoS_2$ and an appearance of several new peaks at $2\theta \approx 30°$, 29° and 27° (Supplementary Fig. S5a) corresponding to decomposition products, such as $K_xS$ [39,49]. Signatures of metallic Mo also appear in XPS spectra (Supplementary Fig. S5b) as a peak at ~228 eV corresponding to Mo $3d_{5/2}$.

The above analysis is summarized in Fig. 4 that maps the observed superconducting phases onto the chemical composition and the structural transitions. The existence of each superconducting phase over a range of average K concentrations corresponds to gradually changing superconducting volume fractions. For example, only 2H-$K_{0.4}MoS_2$ is found for $c \leq 12$ at.%, where parts of the crystals are not intercalated at all (i.e. not superconducting) but the intercalated volume gradually increases with increasing $c$. Only for the average $c \approx 12$ at.% (corresponding to x=0.4 in $K_xMoS_2$) is a complete transition achieved from 2H-$MoS_2$ to 2H-$K_{0.4}MoS_2$ (also see Fig. 3a). For $c > 12$ at.%, atomic coordination in parts of the crystals changes as they transform into 1T-$KMoS_2$, due to local gliding of atomic planes [30] and by $c \approx 25$ at.% (corresponding to x≈1.0 in $K_xMoS_2$) an optimum combination of coexisting 2H and 1T phases is achieved where both phases show maximum contributions to diamagnetic susceptibility (blue symbols in Fig. 1). At the K concentration corresponding to x≈1 ($c \approx 25$ at.%) a further structural transformation occurs where S-Mo-S layers assume a distorted octahedral coordination [31,32,34] and the 1T' superconducting phase (purple symbols in Fig.1) replaces the 1T phase, while still co-existing with the 2H-$K_{0.4}MoS_2$. Finally, at $c > 25$ at.%, the 2H-$K_{0.4}MoS_2$ and 1T'-$KMoS_2$ phases are both present but their volume fractions become quickly reduced as excess K causes disintegration of the $MoS_2$ layers (see above). This is clear from an order of magnitude difference between the diamagnetic moment for $c \approx 25$ at.% (~4·$10^{-2}$ emu/g) and for $c \approx 45$ at.% (~4·$10^{-3}$ emu/g) – see Supplementary Fig. S6. At $c > 45$ at.% superconductivity disappears completely (Supplementary Fig. S6).

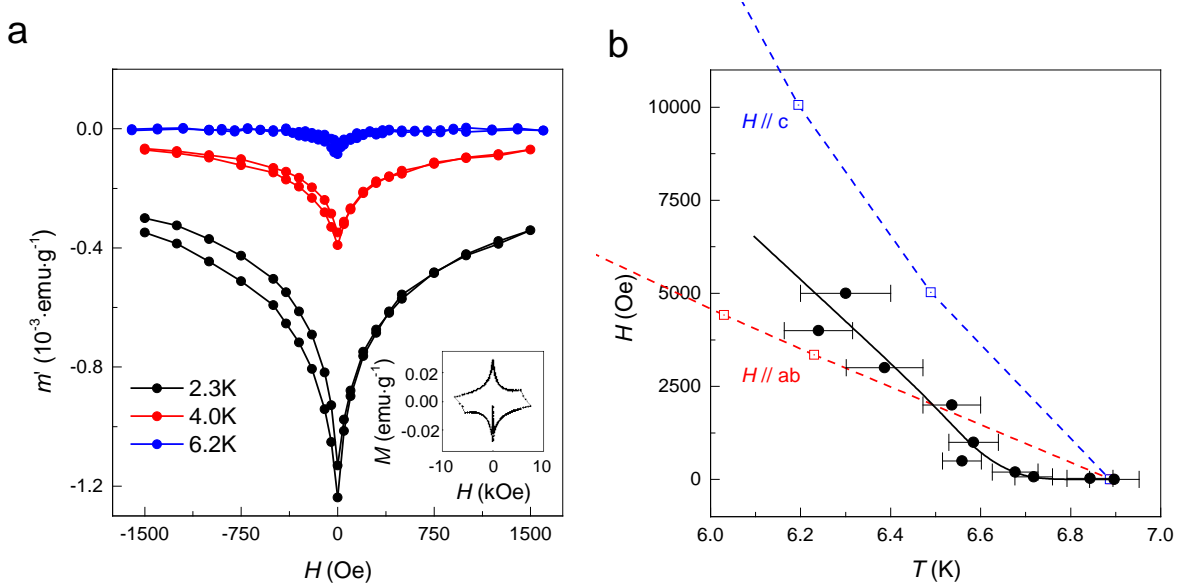

**Figure 5 | Superconducting response of 2H and 1T/1T' $K_xMoS_2$ in magnetic field. (a)** Main panel: ac susceptibility as a function of applied magnetic field for a sample characterized by coexistence of all three superconducting phases ($c \approx 23$ at.% K). The ac drive amplitude is 1 Oe and frequency 80 Hz; ac field is parallel to the dc field. Inset: Typical dc magnetization vs field for the same sample. **(b)** Black symbols: upper critical field, $H_{c2}$, as a function of temperature for the $T_c \approx 6.9K$ phase. Black curve is a guide to the eye. For comparison, blue and red symbols connected by dashed lines show literature data (ref. 20). Due to random orientations of the crystals in our powder samples and strong anisotropy of $H_{c2}$, our data correspond to the average between $H_{c2}^{\parallel}$ and $H_{c2}^{\perp}$. It was not possible to measure $H_{c2}(T)$ for 1T/1T' phases, as explained in the text.

To further characterize the superconducting phases in our samples, we measured the dependence of the diamagnetic moment on the external field, as well as ac susceptibility at temperatures corresponding to the emergence of different phases (i.e. just below the respective $T_c$'s). Typical characteristics are shown for a sample with $c \approx 25$ at.% where all three superconducting phases coexist (Fig. 5): ac susceptibilities for all three phases show behaviour typical for type-II superconductors (Fig. 5a). Furthermore, $H_{c2}(T)$ for the highest $T_c$ phase (2H) exhibits a pronounced positive curvature (Fig. 5b), in agreement with earlier studies [20,50]. We note that the measurements of $T_c$ in a meaningful range of $H$ could only be done for the 2H-$K_{0.4}MoS_2$ phase that has the highest $T_c$. For the 1T-$KMoS_2$ ($T_c \approx 2.8$K) $T_c$ becomes quickly suppressed by the applied magnetic field to below the lowest available temperature of the experiments (1.8K), while for the 1T'-$KMoS_2$ phase ($T_c \approx 4.6$K) the measurements are complicated by the emergence of a strong paramagnetic signal from clusters of excess potassium [51,52], which makes it difficult to accurately determine the temperature corresponding to the onset of the diamagnetic signal.

The positive curvature in $H_{c2}(T)$ for alkali-metal intercalated $MoS_2$ was noticed already in the 1970s and its origin discussed intensively at the time. Later a similar positive curvature was found for many other layered compounds – other intercalated TMDs, organic superconductors, high-temperature superconductors and artificial multilayers made of alternating superconductor-normal metal or superconductor-insulator layers. The origin of the positive curvature is now understood as a signature of a dimensional crossover: A common characteristic of all these materials is that they are made up of weakly-coupled superconducting layers [53]: At temperatures just below $T_c$, the superconducting coherence length in the direction perpendicular to the layers, $\xi_\perp$, is much larger than the interlayer separation and the coupling between the layers relatively strong, so that the material behaves as a 3D superconductor, with a linear slope of $H_{c2}(T)$. As the temperature is lowered, $\xi_\perp$ decreases, the coupling between the layers becomes Josephson-like and the response to the parallel magnetic field becomes similar to that of a thin (2D) film with a negligibly small thickness. As such 2D films in parallel magnetic fields are characterized by a diverging slope of $H_{c2}(T)$ at temperatures just below $T_c$ (ref. 54), for a layered superconductor such $K_xMoS_2$ the crossover from 3D to 2D behaviour results in a pronounced positive curvature, as observed for all alkali-metal intercalated $MoS_2$ in early studies, and also for the 2H-$K_{0.4}MoS_2$ phase in our experiment.

It is instructive to compare the reported superconducting phases of metallic $MoS_2$ with recent findings of superconductivity in other metallic TMD polymorphs, in particular 1T-$TaS_2$ and 1T-$TiSe_2$ [12,15,16,21-24]. The superconducting state in those studies was induced either by chemical doping (intercalation) [22], high pressure [15,16] or intercalation combined with electrostatic doping by ionic gating [12]. A common feature of the superconducting state in the above TMDs is that it emerges from one of the charge-density-wave (CDW) states [21-25], with microscopic/mesoscopic-scale phase separation being an essential feature of the electronic states preceding the emergence of superconductivity (see e.g. ref. 23). As both electronic states – CDWs and superconductivity – are underpinned by strong electron-phonon interactions, there has been much effort to understand the relation between them. An emergence of CDW ordering was also predicted for highly doped $MoS_2$ [26], although the possibility of doping-induced transition to the metallic 1T/1T' phases (such as observed in our work and shown in several TEM studies) was not taken into account. According to first-principle calculations in ref. 26, the superconducting state is expected to emerge first with increasing doping and then replaced by the CDW phase, i.e., a reversed order of phase transitions compared to doped $TaS_2$ and $TiSe_2$. In our experiments, we could not find any signatures of CDWs even at the highest doping levels; such signatures can in principle be seen in $\chi(T)$ as shown in e.g. ref. 22. It is possible that CDW-induced changes in susceptibility of doped $MoS_2$ are more subtle that in $Cu_xTiSe_2$ (ref. 22) and were not noticeable in our measurements. Another possibility is that CDW state in $MoS_2$ requires even higher K concentrations than we were able to achieve or, alternatively, that $MoS_2$ is different from the above family of TMDs and the CDW phase is absent, as suggested in ref. 13.

The electronic density responsible for superconductivity in 1T and 1T' phases in our experiments is determined by a combination of charge transfer from K atoms and a finite electron density in pristine metallic 1T/1T' phase. Experimentally it is not possible to separate these two contributions as the transition to the metallic state is itself caused by doping. Furthermore, it was not possible to directly measure the carrier concentrations, $n$, corresponding to different superconducting phases in our case, due to the fact that different phases were always mixed in the same crystals. Nevertheless we can estimate the lower limit on $n$ by analogy with intercalation-induced doping of the semiconducting polymorph 2H-$MoS_2$. In the latter case, $n$ corresponding to the stoichiometric compound $K_{0.4}MoS_2$ was found to be $\approx 1.1 \cdot 10^{22}$ cm$^{-3}$ (ref. 55). If we assume the same rate of electron transfer from K atoms as their concentration increases by 2.5 times to yield $KMoS_2$ (i.e., the composition corresponding to the 2H-1T' phase transition), the electron concentration in the superconducting metallic 1T/1T'-$KMoS_2$ must be $>2.7 \cdot 10^{22}$ cm$^{-3}$. This is notably higher than the reported $n \approx 1.1 \cdot 10^{22}$ cm$^{-3}$ needed to achieve superconductivity in 1T-$TaS_2$ [12]. Another indication that achieving a superconducting state in metallic $MoS_2$ requires significantly higher electron concentrations comes from experiments where the metallic state and superconductivity were induced in pristine 1T-$TaS_2$ and 2H-$MoS_2$ by ultra-high pressure (refs. 15 and 13, respectively). Pressure in these experiments was used as a tool to influence the electronic interactions and cause transitions to the metallic and ultimately superconducting state, in a similar way to the effect of intercalation. While pressures $P$ of the order of 2-4 GPa were sufficient to induce superconductivity in 1T-$TaS_2$ [15], achieving it in pristine $MoS_2$ requires $P$ in excess of 90 GPa, according to a recent report [13] and $P \sim 20$ GPa was needed just to close the bandgap and bring $MoS_2$ into a metallic state [56].

Finally, we note that the well separated superconducting transitions in our experiments are an indication of a coexistence of locally ordered and sharply defined structural phases. This is in agreement with atomic-resolution TEM studies that found that the boundaries between 2H and 1T/1T' phases are always atomically sharp with no visible defects[32,34,35]. This creates opportunities to use $MoS_2$ for nanoscale engineering of superconducting devices, for example, Josephson junctions involving different superconducting phases or a superconducting and a semiconducting or metallic phase that can be locally tuned by doping.

## Methods

Intercalation was carried out in a quartz reactor tube that could be evacuated and filled with ammonia or other gases as required. The starting materials [pristine 2H-$MoS_2$ powder (Aldrich 99%) and K metal (Aldrich 99.95%)] were sealed inside the reactor tube in the inert atmosphere of a glove box, where oxygen and moisture levels were maintained at <0.5 ppm. After evacuating the reactor to $\sim 10^{-5}$ mbar, it was placed in a bath of dry ice/ethanol (temperature $T \approx -78$ °C) and filled with high-purity ammonia gas (CK Gas 99.98%). The latter condensed onto the reactants, forming a deep-blue solution of K in liquid ammonia. After that the reactor was kept in dry ice/ethanol bath for 24 h – 500 h, depending on the desired degree of intercalation. To prevent oxygen or moisture from the surrounding air entering the reactor during intercalation, the space formed after ammonia condensation was filled with high-purity argon gas. To recover the product, ammonia was pumped out of the reactor and the intercalated crystals taken out in the inert atmosphere of a glove box. Due to extreme sensitivity of alkali-metal intercalated samples to oxygen and moisture, they were either handled in the inert atmosphere of a glove box or protected by immersion in paraffin oil or by using special sealed containers.

Energy-dispersive X-ray spectroscopy (EDS) was carried out using Oxford Instruments X-Max detector integrated with the scanning electron microscope Zeiss Ultra Field Emission SEM. The samples were prepared and placed in a sealed container inside the glovebox, then quickly transferred into the load-lock chamber of the SEM. To determine K content for each intercalation run, EDS measurements were carried out at about ten different randomly selected points on each sample and the average taken as the measured K concentration, $c$.

X-ray powder diffraction (XRD) measurements were performed using Bruker D8 Discover diffractometer with Cu $K\alpha$ radiation ($\lambda$ = 1.5406 Å). Before each measurement, intercalated crystals were ground and mixed with paraffin oil, then sealed in an air-tight XRD sample holder (Bruker, A100B36/B37). X-ray diffraction spectra were obtained at room temperature in the $2\theta$ range 5° - 70°, with a step of 0.03° and a time step of 0.5 s.

To obtain high-resolution X-ray photoelectron spectra (XPS) we used Kratos Axis Ultra spectrometer. All sample preparation was done in a glove box connected to the spectrometer's load-lock chamber. Intercalated crystals (typical dimensions 10 µm × 1 µm) were carefully sprinkled on a double-sided adhesive tape with an area of 3 mm × 3 mm. In this way 15 samples could be prepared at the same time on a sample holder with an available area of 8 cm × 1.5 cm. For each sample spectra were taken from an area of 700 microns to 300 microns in diameter, using an Al $K\alpha$ micro-focused monochromatized source with a step of 0.1 eV and dwell time of 300 ms. All spectra were calibrated using Carbon 1s peak located at 284.8 eV [47].

Magnetization measurements were carried out using Quantum Design MPMS-XL7 SQUID magnetometer. To prevent degradation of the intercalated samples during transfer to the cryostat and subsequent measurements, they were immersed in paraffin oil and sealed inside polycarbonate capsules in dry argon atmosphere of a glove box, then quickly transferred to the cryostat and immediately cooled down to below ~50K. To check that these precautions were sufficient to ensure sample stability, we repeated several measurements 1-2 months after the samples were prepared and measured for the first time. This showed highly reproducible results: an example is shown in Supplementary Fig. S7. There was no detectable change in magnetization behaviour after a month of keeping the sample in a glove box.

For zero-field-cooling (ZFC) measurements a ~10mg sample of intercalated crystals was cooled to 1.8K in zero applied field, then an external field applied at 1.8K and the magnetic moment measured as the temperature increased to 15-20K. For field-cooling (FC) measurements a magnetic field $H$ was applied at a temperature well above the superconducting transition, typically ~15 K, and the magnetic moment measured as the sample was cooled down to 1.8K. The upper critical field, $H_{c2}(T)$, in Fig. 5 was extracted from measurements of $H$-dependent critical temperature, $T_c$. Due to the very small size of individual crystals they were used as powder, so that it was not possible to control the magnetic field orientation with respect to the crystal axes. As $K_xMoS_2$ is strongly anisotropic, the measured $H_{c2}(T)$ values represent the average between $H_{c2}^{\parallel}$ and $H_{c2}^{\perp}$. AC susceptibility was measured with 1 Oe, 80 Hz ac field parallel to the dc field.

# References

(1) Xi, X.; Zhao, L.; Wang, Z.; Berger, H.; Forró, L.; Shan, J.; Mak, K. F. *Nat. Nanotechnol.* **2015,** 10, 765-770.
(2) Cao, Y.; Mishchenko. A.; Yu, G. L.; Khestanova, E.; Rooney, A. P.; Prestat, E.; Kretinin, A. V.; Blake, P.; Shalom, M. B.; Woods, C.; Chapman, J.; Balakrishnan, G.; Grigorieva, I. V.; Novoselov, K. S.; Piot, B. A.; Potemski, M.; Watanabe, K.; Taniguchi, T.; Haigh, S. J.; Geim, A. K.; Gorbachev, R. V. *Nano Lett.* **2015,** 15, 4914–4921.
(3) Staley, N. E.; Wu, J.; Eklund, P.; Liu, Y.; Li, L.; Xu, Z. *Phys. Rev. B,* **2009,** 80 (18), 184505.
(4) Ge, J. F.; Liu, Z. L.; Liu, C.; Gao, C. L.; Qian, D.; Xue, Q. K.; Liu, Y.; Jia, J. F. *Nature Mater.* **2015,** 14, 285-289.
(5) Friend, R. H.; Yoffe, A. D. *Adv. Phys.* **1987,** 36 (1), 1-94.
(6) Weller, T. E.; Ellerby, M.; Saxena, S. S.; Smith, R. P.; Skipper, N. T. *Nature Phys.* **2005,** 1, 39-41.
(7) Wray, L. A.; Xu, S. Y.; Xia, Y.; San Hor, Y.; Qian, D.; Fedorov, A. V.; Lin, H.; Bansil, A.; Cava, R. J.; Hasan, M. Z. *Nature Phys.* **2010,** 6 (11), 855-859.
(8) Burrard-Lucas, M.; Free, D. G.; Sedlmaier, S. J.; Wright, J. D.; Cassidy, S. J.; Hara, Y.; Corkett, A. J.; Lancaster, T.; Baker, P. J.; Blundell, S. J.; Clarke, S. J. *Nature Mater.* **2013,** 12 (1), 15-19.
(9) Ye, J. T.; Zhang, Y. J.; Akashi, R.; Bahramy, M. S.; Arita, R.; Iwasa, Y. *Science* **2012,** 338 (6111), 1193-1196.
(10) Taniguchi, K.; Matsumoto, A.; Shimotani, H.; Takagi, H. *Appl. Phys. Lett.* **2012,** 101 (4), 042603.
(11) Jo, S.; Costanzo, D.; Berger, H.; Morpurgo, A. F. *Nano Lett.* **2015,** 15 (2), 1197-1202.
(12) Yu, Y.; Yang, F.; Lu, X. F.; Yan, Y. J.; Cho, Y. H.; Ma, L.; Niu, X.; Kim, S.; Son, Y. W.; Feng, D.; Li, S.; Cheong, S. W.; Chen, X. H.; Zhang, Y. *Nature Nanotechnol.* **2015,** 10 (3), 270-276.
(13) Chi, Z.; Yen, F.; Peng, F.; Zhu, J.; Zhang, Y.; Chen, X.; Yang, Z.; Liu, X.; Ma, Y.; Zhao, Y.; Kagayama, T.; Iwasa, Y. *arXiv preprint*. **2015,** arXiv*:1503.05331*.
(14) Pan, X. C.; Chen, X.; Liu, H.; Feng, Y.; Wei, Z.; Zhou, Y.; Chi, Z.; Pi, L.; Yen, F.; Song, F.; Wan, X.; Yang, Z.; Wang, B.; Wang, G.; Zhang, Y. *Nature Commun.* **2015,** 6, 7805.
(15) Sipos, B.; Kusmartseva, A. F.; Akrap, A., Berger, H.; Forró, L.; Tutiš, E. *Nature Mater.* **2008,** 7 (12), 960-965.
(16) Kusmartseva, A. F.; Sipos, B.; Berger, H.; Forró, L.; Tutiš, E. *Phys. Rev. Lett.* **2009,** 103 (23), 236401.
(17) Somoano, R. B.; Rembaum, A. *Phys. Rev. Lett.* **1971,** 27 (7), 402.
(18) Somoano, R. B.; Hadek, V.; Rembaum, A. *J. Chem. Phys.* **1973,** 58 (2), 697-701.
(19) Somoano, R. B.; Hadek, V.; Rembaum, A.; Samson, S.; Woollam, J. *J. Chem. Phys.* **1975,** 62 (3), 1068-1073.
(20) Woollam, J. A.; Somoano, R. B. *Phys. Rev.* B **1976,** 13 (9), 3843.
(21) Wilson, J. A.; Di Salvo, F. J.; Mahajan, S. *Adv. Phys.* **1975,** 24 (2), 117-201.
(22) Morosan, E.; Zandbergen, H. W.; Dennis, B. S.; Bos, J. W. G.; Onose, Y.; Klimczuk, T.; Ramirez, A. P.; Ong, N. P.; Cava, R. J. *Nature Phys.* **2006,** 2 (8), 544-550.
(23) Joe, Y. I.; Chen, X. M.; Ghaemi, P.; Finkelstein, K. D.; de La Peña, G. A.; Gan, Y.; Lee, J. C.; Yuan, S.; Geck, J. MacDougail, G. J.; Chiang, T. C.; Cooper, S. L.; Fradkin, E.; Abbamonte, P. *Nature Phys.* **2014,** 10 (6), 421-425.
(24) Calandra, M.; Mauri, F. *Phys. Rev. Lett.* **2011,** 106 (19), 196406.
(25) Ang, R.; Tanaka, Y.; Ieki, E.; Nakayama, K.; Sato, T.; Li, L. J.; Lu, W. J.; Sun, Y. P.; Takahashi, T. *Phys. Rev. Lett.* **2012,** 109 (17), 176403.
(26) Rösner, M.; Haas, S.; Wehling, T. O. *Phys. Rev. B.* **2014,** 90 (24), 245105.
(27) Roldán, R.; Cappelluti, E.; Guinea, F. *Phys. Rev. B.* **2013,** 88 (5), 054515.

(28) Yuan, N. F. Q.; Mak, K. F.; Law, K. *Phys. Rev. Lett.* **2014,** 113 (9), 097001.
(29) Wypych, F.; Schöllhorn, R. *J.Chem. Soc. Chem. Commun.* **1992,** 19, 1386–1388.
(30) Lin, Y. C.; Dumcenco, D. O.; Huang, Y. S.; Suenaga, K. *Nature Nanotechnol.* **2014,** 9 (5), 391-396.
(31) Duerloo, K. A. N.; Li, Y.; Reed, E. J. *Nature Commun.* **2014,** 5, 4214.
(32) Eda, G.; Fujita, T.; Yamaguchi, H.; Voiry, D.; Chen, M.; Chhowalla, M. *ACS Nano.* **2012,** 6 (8), 7311-7317.
(33) Chhowalla, M.; Shin, H. S.; Eda, G.; Li, L. J.; Loh, K. P.; Zhang, H. *Nature Chem.* **2013,** 5 (4), 263-275.
(34) Cheng, Y.; Nie, A.; Zhang, Q.; Gan, L. Y.; Shahbazian-Yassar, R.; Schwingenschlogl, U. *ACS Nano.* **2014,** 8 (11), 11447-11453.
(35) Kappera, R.; Voiry, D.; Yalcin, S. E.; Branch, B.; Gupta, G.; Mohite, A. D.; Chhowalla, M. *Nature Mater.* **2014,** 13 (12), 1128-1134.
(36) Voiry, D.; Goswami, A.; Kappera, R.; e Silva, C. D. C. C.; Kaplan, D.; Fujita, T.; Chen, M.; Asefa, T.; Chhowalla, M. *Nature Chem.* **2015,** 7 (1), 45-49.
(37) Py, M. A.; Haering, R. R. *Can. J. Phys.* **1983,** 61 (1), 76-84.
(38) Acerce, M.; Voiry, D.; Chhowalla, M. *Nature Nanotechnol.* **2015,** 10 (4), 313-318.
(39) Wang, X.; Shen, X.; Wang, Z.; Yu, R.; Chen, L. *ACS Nano.* **2014,** 8 (11), 11394-11400.
(40) Gao, P.; Wang, L.; Zhang, Y.; Huang, Y.; Liu, K. *ACS Nano.* **2015**, DOI: 10.1021/acsnano.5b04950.
(41) Enyashin, A. N.; Yadgarov, L.; Houben, L.; Popov, I.; Weidenbach, M.; Tenne, R.; Bar-Sadan, M.; Seifert, G. *J. Phys. Chem. C.* **2011,** 115 (50), 24586-24591.
(42) Kang, Y.; Najmaei, S.; Liu, Z.; Bao, Y.; Wang, Y.; Zhu, X.; Halas, N. J.; Nordlander, P.; Ajayan, P. M.; Lou, J.; Fang, Z. *Adv. Mater.* **2014,** 26 (37), 6467-6471.
(43) Kan, M.; Wang, J. Y.; Li, X. W.; Zhang, S. H.; Li, Y. W.; Kawazoe, Y.; Sun, Q.; Jena, P. *J. Phys. Chem. C.* **2014,** 118 (3), 1515-1522.
(44) Eda, G.; Yamaguchi, H.; Voiry, D.; Fujita, T.; Chen, M.; Chhowalla, M. *Nano Lett.* **2011,** 11 (12), 5111-5116.
(45) Wang, H.; Lu, Z.; Xu, S.; Kong, D.; Cha, J. J.; Zheng, G.; Hsu, P. C.; Yan, K.; Bradshaw, D.; Prinz, F. B.; Cui, Y. *PNAS.* **2013,** 110 (49), 19701-19706.
(46) Catterall, R.; Mott, N. F. *Adv. Phys.* **1969,** 18 (76), 665-680.
(47) Wagner, C. D. *Handbook of X-ray Photoelectron Spectroscopy: A Reference Book of Standand Data for Use in X-ray Photoelectron Spectroscopy.* Perkin-Elmer Corp. 1979.
(48) Kertesz, M.; Hoffmann, R. *J. Am. Chem. Soc.* **1984,** 106 (12), 3453-3460.
(49) Wang, L.; Xu, Z.; Wang, W.; Bai, X. *J. Am. Chem. Soc.* **2014,** 136 (18), 6693-6697.
(50) Woollam, J. A.; Somoano, R. B.; O'Connor, P. *Phys. Rev. Lett.* **1974,** 32 (13), 712.
(51) Rocquefelte, X.; Boucher, F.; Gressier, P.; Ouvrard, G.; Blaha, P.; Schwarz, K. *Phys. Rev. B.* **2000,** 62 (4), 2397.
(52) Takai, K.; Eto, S.; Inaguma, M.; Enoki, T.; Ogata, H.; Tokita, M.; Watanabe, J. *Phys. Rev. Lett.* **2007,** 98 (1), 017203.
(53) Klemm, R. A. *Layered Superconductors.* Volume 1, chapter 7. Oxford University Press, 2012.
(54) M. Tinkham. *Phys. Rev. B* **1963**, 129 (6), 2413-2422.
(55) Woollam, J. A.; Somoano, R. B. *Mater. Sci. Eng.* **1977,** 31, 289-295.
(56) Nayak, A. P.; Bhattacharyya, S.; Zhu, J.; Liu, J.; Wu, X.; Pandey, T.; Jin, C.; Singh, A. K.; Akinwande, D.; Lin, J. F. *Nature Commun.* **2014,** 5, 3731.

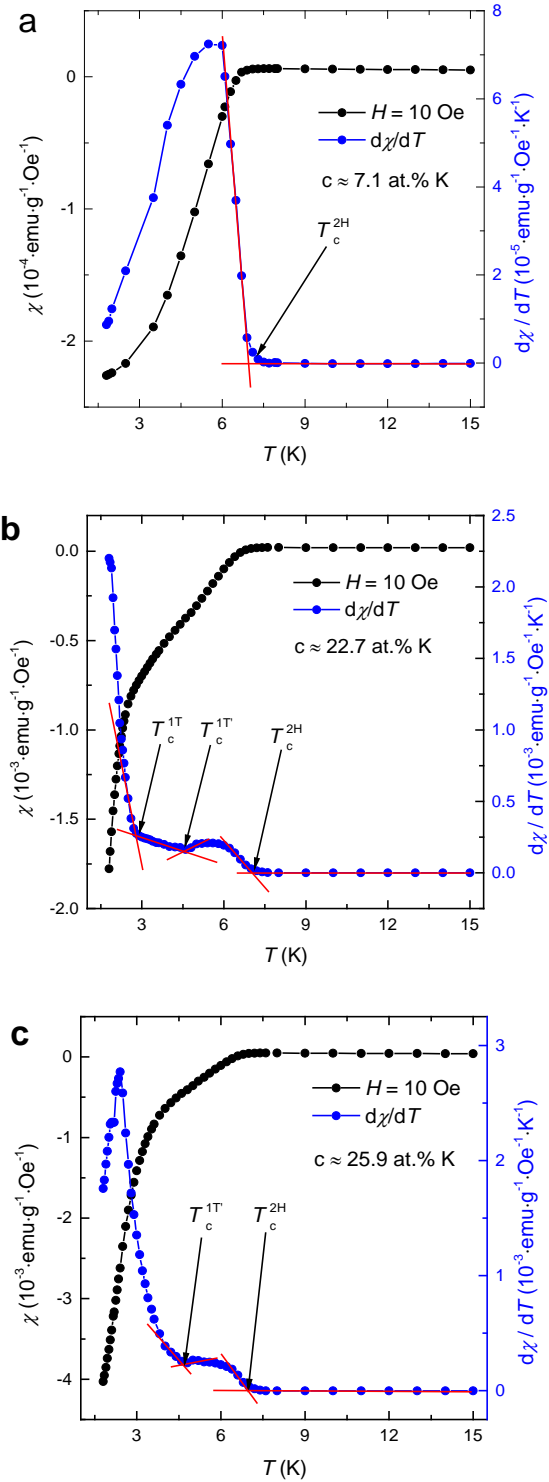

**Supplementary Figure S1 | Determination of the critical temperatures, $T_c$.** Black curves in (a), (b) and (c) are temperature dependent dc susceptibilities, $\chi(T) = M/H$, for $c \approx 7$ at.%, 23 at.% and 26 at.% K, respectively. Blue curves show the corresponding numerical derivatives $d\chi/dT$. Each $T_c$ is determined as the temperature corresponding to a sharp increase in $d\chi/dT$ and is given by a cross-point of linear extrapolations of the corresponding parts of the curve.

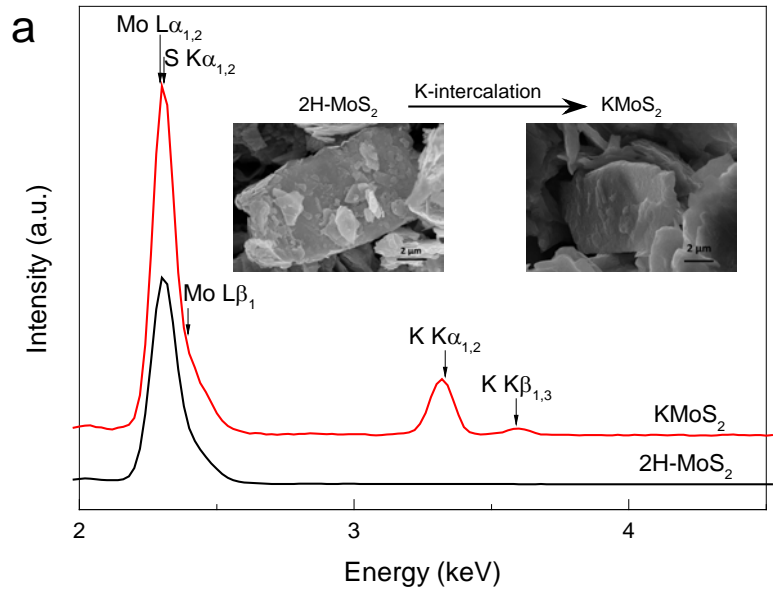

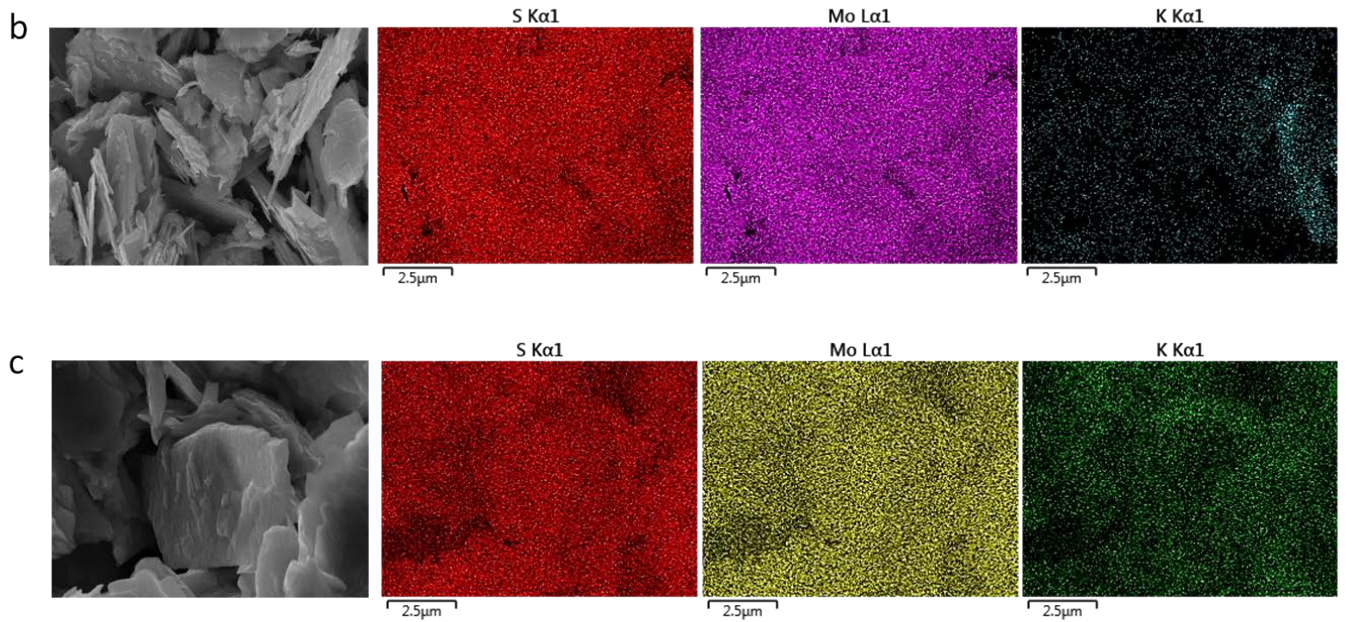

**Supplementary Figure S2 | Distribution and concentrations of K in intercalated MoS$_2$ from EDS analysis. (a)** Typical EDS spectra obtained from pristine (2H-MoS$_2$) and intercalated (KMoS$_2$) crystals such as those shown as insets. **(b,c)** Elemental EDS maps and corresponding SEM images for samples with $c \approx 7$ at.% K and 23 at.%, respectively. Distribution of K after 12 hours of intercalation ($c \approx 7$ at.% K) is not uniform, in agreement with very small superconducting fraction for this sample (see main text). For the fully intercalated sample (intercalation time ~100 hours) shown in (c), the distribution of K (K$\alpha$1) follows closely the shape of the crystals as seen from the distribution of Mo and S atoms (Mo L$\alpha$1 and S K$\alpha$1 maps, respectively).

| Sample number | S1 | S2 | S3 | S4 | S5 | S6 | S7 | S8 | S9 | S10 | S11 |
|---|---|---|---|---|---|---|---|---|---|---|---|
| Average K concentration, $c$ (at.%) | 4.0 ± 0.6 | 7.1 ± 0.3 | 9.3 ± 0.6 | 11.4 ± 0.6 | 15.2 ± 0.9 | 17.6 ± 0.6 | 22.7 ± 0.3 | 25.9 ± 1.9 | 36.5 ± 3.5 | 41.4 ± 3.8 | 47.0 ± 5.1 |

**Supplementary Table S1 | Average K concentration in intercalated MoS$_2$.** Different samples correspond to different intercalation times (intercalation time increases from sample S1 to sample S11). For each sample K concentration was measured at 10 randomly selected points and the average used as parameter $c$ (see main text). Large standard derivations for sample S9 – S11 are due to clustering of potassium as explained in the main text.

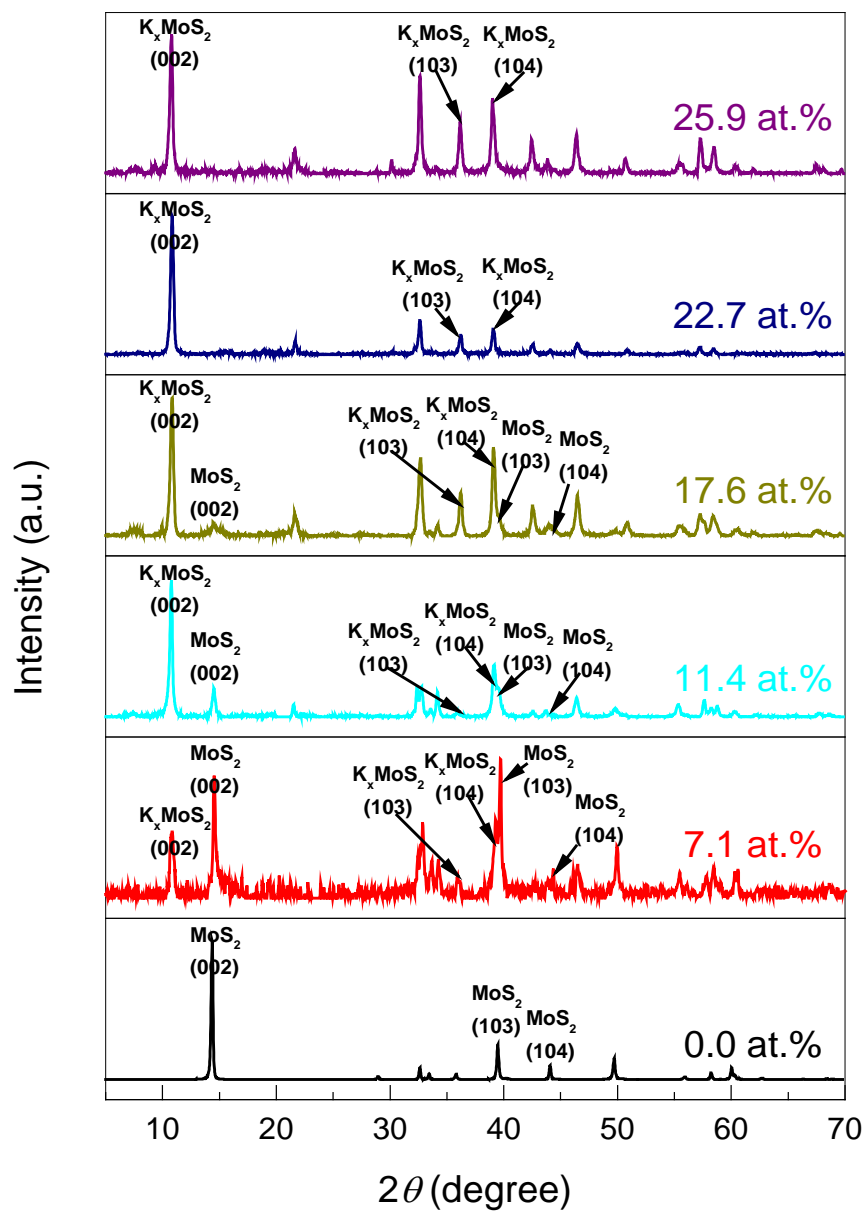

**Supplementary Figure S3 | Labelling of XRD peaks.** Examples of XRD spectra for pristine (bottom curve) and intercalated MoS$_2$, at different average K concentrations. Labels are shown for all main peaks. Average K concentrations, *c*, are shown by numbers above each spectrum. The shifts of (*002*), (*103*) and (*104*) peaks to lower 2θ indicate the expansion of the crystal lattice along the *c*-axis upon intercalation.

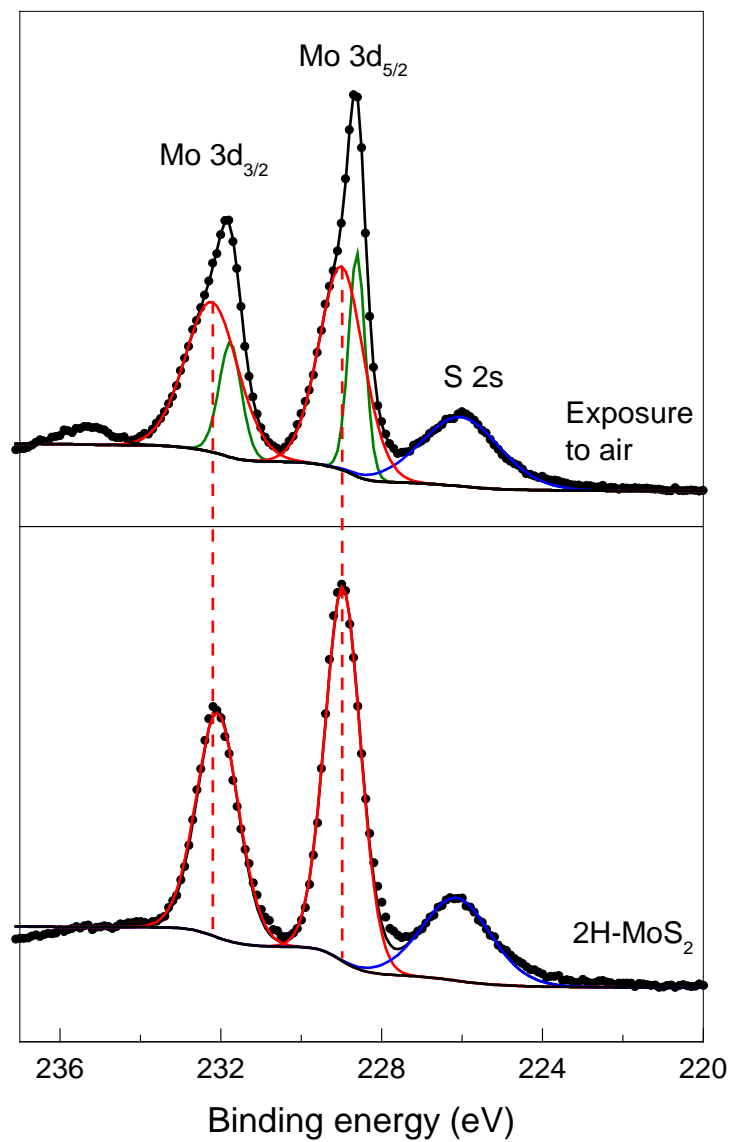

**Supplementary Figure S4 | XPS characterization of a $K_xMoS_2$ sample after exposure to air.** Bottom panel: pristine sample (before intercalation); top panel: after intercalation to $c \approx 20\%$ and exposure to air. Measured XPS spectra are shown in black; red and green curves correspond to 2H and 1T phase, respectively. De-intercalation of K atoms is clear from the shift of Mo peaks corresponding to the 2H phase to 'pristine' positions while the peaks corresponding to 1T/1T' phase remain.

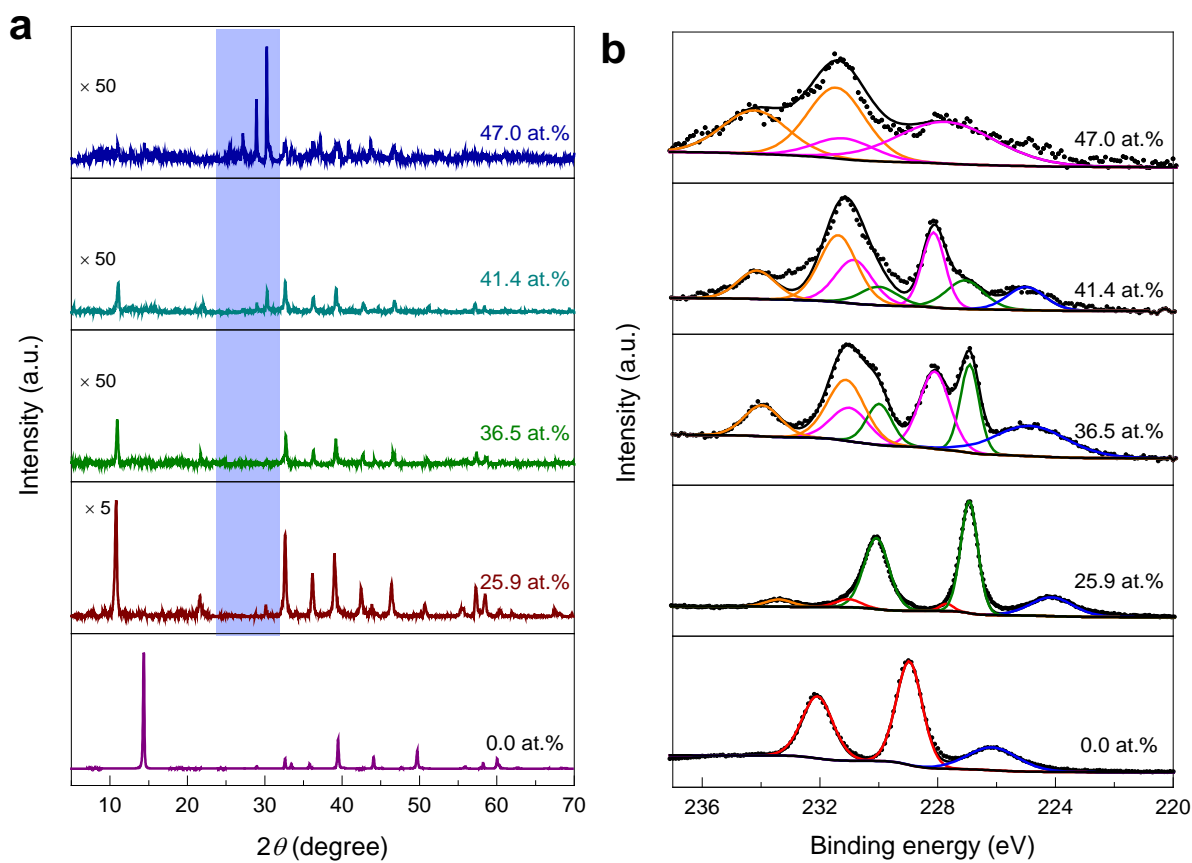

**Supplementary Figure S5 | Characterization of the decomposition reaction at high potassium concentrations. (a)** XRD and **(b)** XPS spectra for KMoS$_2$ samples with average K concentrations in excess of 25 at.%. Bottom curves are spectra for pristine 2H-MoS$_2$ shown for comparison. The average K concentrations are shown as numerical labels above each spectrum. New peaks in XRD spectra corresponding to K$_x$S (similar to Na$_x$S and Li$_x$S reported in refs. [39,49]) appear at $c$≥26 at.% (highlighted in blue). Due to the quickly decreasing intensity of XRD peaks for $c$ ≥26 at.%, each spectrum is multiplied by a numerical factor shown on the left. The measured XPS spectra in (b) are shown in black; red, green, pink and orange curves correspond to the 2H phase, 1T phase, metallic Mo and Mo oxide components, respectively [39,47,49]. At $c$ ≈26 at.% new Mo 3d peaks appear at higher binding energies, indicating increasing oxidation of Mo. At higher still $c$, these peaks increase in intensity, new peaks appear and all are eventually replaced by peaks corresponding to metallic Mo (pink line) and molybdenum oxide (orange line).

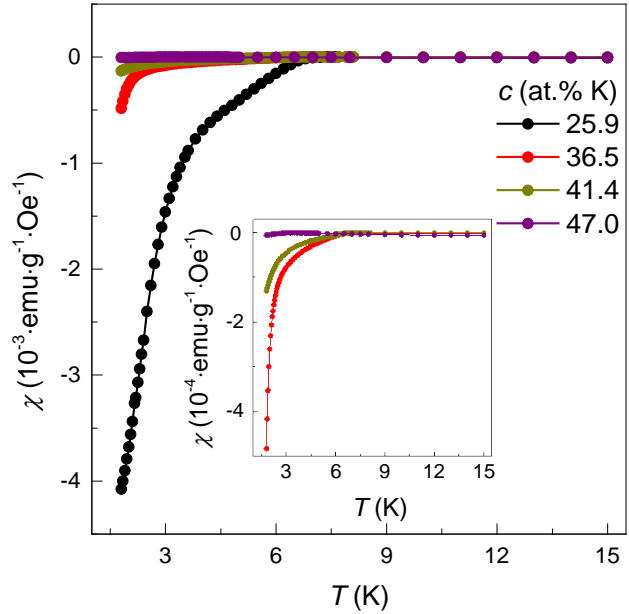

**Supplementary Figure S6 | Disappearance of superconductivity at high K concentrations.** Shown is the temperature-dependent magnetic susceptibility, $\chi = M/H$, for intercalated samples with $c > 26$ at.%. Here $M$ is the magnetic moment measured in ZFC mode at $H = 10$ Oe. The rapid decrease in $\chi$ by an order of magnitude for $c \approx 37$ at.% and the complete disappearance of the diamagnetic response at $c \approx 47$ at.% are due to sample decomposition (see main text).

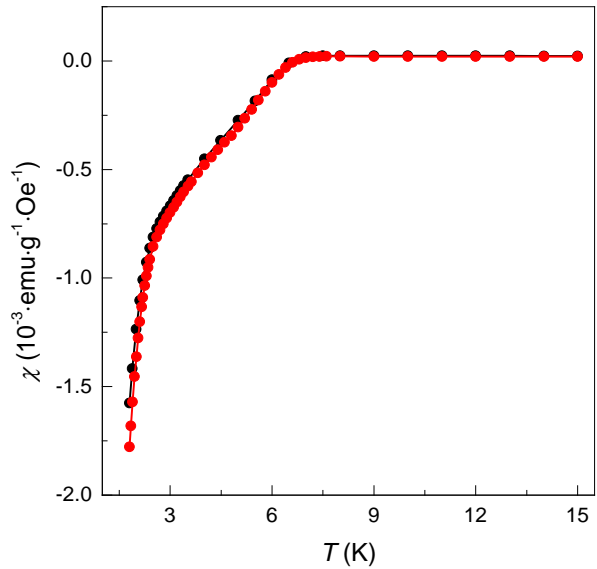

**Supplementary Figure S7 | Reproducibility of the superconducting response and stability of the superconducting phases.** Shown is the temperature-dependent dc magnetic susceptibility, $\chi = M/H$, for an intercalated sample with $c \approx 23$ at.% measured immediately after intercalation (black symbols) and after one months in the glove box (red symbols). Here $M$ is the magnetic moment measured in ZFC mode at $H = 10$ Oe.